\documentclass[12pt]{article}
\usepackage[centertags]{amsmath}
\usepackage{amssymb}
\usepackage{amsthm}
\usepackage[dvips]{graphicx}
\usepackage{mathrsfs}
\usepackage{amsfonts}
\usepackage{color}
\usepackage{colordvi,bm}
\usepackage{hyperref}

\newcommand{\be}{\begin{equation}}
\newcommand{\ee}{\end{equation}}
\newcommand{\bea}{\begin{eqnarray}}
\newcommand{\eea}{\end{eqnarray}}
\newcommand{\bmx}{\begin{pmatrix}}
\newcommand{\emx}{\end{pmatrix}}
\newcommand{\sox}{\otimes_s}

\begin{document}

\vfuzz2pt % Don't report over-full v-boxes if over-edge is small
\hfuzz2pt % Don't report over-full h-boxes if over-edge is small

%----- Load Definitions ------

% THEOREMS -------------------------------------------------------
\theoremstyle{definition}
\newtheorem{thm}{Theorem}[section]
\newtheorem{cor}[thm]{Corollary}
\newtheorem{lem}[thm]{Lemma}
\newtheorem{prop}[thm]{Proposition}
\theoremstyle{definition}
\newtheorem{defn}[thm]{Definition}
\theoremstyle{definition}
\newtheorem{fact}[thm]{Fact}
\theoremstyle{definition}
\newtheorem{rem}[thm]{Remark}
\numberwithin{equation}{section}

% MATH -----------------------------------------------------------

\newcommand{\nn}{\nonumber}

\newcommand{\delslash}{\hbox{\ooalign{$\displaystyle {\nabla}$\cr$\hspace{.02in}/$}}}
\newcommand{\pdslash}{\hbox{\ooalign{$\displaystyle {\partial}$\cr$\hspace{0in}/$}}}

\newcommand{\Dslash}{\hbox{\ooalign{$\displaystyle {D}$\cr$\hspace{.02in}/$}}}
\newcommand{\Pslash}{\hbox{\ooalign{$\displaystyle {P}$\cr$\hspace{.02in}/$}}}

\newcommand{\half}{\frac{1}{2}}
\newcommand{\third}{\frac{1}{3}}
\newcommand{\fourth}{\frac{1}{4}}

\newcommand{\ihalf}{{i\over 2}}

\newcommand{\Z}{\mathbb{Z}}
\newcommand{\R}{\mathbb{R}}
\newcommand{\C}{\mathbb{C}}
\newcommand{\Hq}{\mathbb{H}}
\newcommand{\gr}[1]{{\left| #1 \right|}}
\newcommand{\Pb}{\mathbf{P}}
\newcommand{\Tb}{\mathbf{T}}

\newcommand{\abf}{\mathbf{a}}
\newcommand{\bbf}{\mathbf{b}}
\newcommand{\ebf}{\mathbf{e}}
\newcommand{\fbf}{\mathbf{f}}
\newcommand{\gbf}{\mathbf{g}}
\newcommand{\hbf}{\mathbf{h}}
\newcommand{\ibf}{\mathbf{i}}
\newcommand{\jbf}{\mathbf{j}}
\newcommand{\pbf}{\mathbf{p}}
\newcommand{\tbf}{\mathbf{t}}
\newcommand{\vbf}{\mathbf{v}}
\newcommand{\wbf}{\mathbf{w}}
\newcommand{\xbf}{\mathbf{x}}
\newcommand{\ybf}{\mathbf{y}}
\newcommand{\zbf}{\mathbf{z}}

\newcommand{\Abf}{\mathbf{A}}
\newcommand{\Bbf}{\mathbf{B}}
\newcommand{\Ebf}{\mathbf{E}}
\newcommand{\Fbf}{\mathbf{F}}
\newcommand{\Gbf}{\mathbf{G}}
\newcommand{\Hbf}{\mathbf{H}}
\newcommand{\Ibf}{\mathbf{I}}
\newcommand{\IIbf}{\mathbf{II}}
\newcommand{\IIIIbf}{\mathbf{III}}
\newcommand{\Jbf}{\mathbf{J}}
\newcommand{\Pbf}{\mathbf{P}}
\newcommand{\Tbf}{\mathbf{T}}
\newcommand{\Vbf}{\mathbf{V}}
\newcommand{\Wbf}{\mathbf{W}}
\newcommand{\Xbf}{\mathbf{X}}
\newcommand{\Ybf}{\mathbf{Y}}
\newcommand{\Zbf}{\mathbf{Z}}

\newcommand{\TwoMink}{\mathbb{R}^{1,1}}
\newcommand{\TwoHMink}{\mathbb{H}^{1,1}}
\newcommand{\Ppoint}{(\mathscr{P})}
\newcommand{\DMink}{\mathbb{R}^{1,D-1}}
\newcommand{\SOC}{SO^\uparrow}

\renewcommand{\sl}{\mathfrak{sl}}
\newcommand{\so}{\mathfrak{so}}
\newcommand{\su}{{\mathfrak{su}}}
\newcommand{\iso}{\mathfrak{iso}}

\newcommand{\afrak}{\mathfrak{a}}
\newcommand{\bfrak}{\mathfrak{b}}
\newcommand{\cfrak}{\mathfrak{c}}
\newcommand{\dfrak}{\mathfrak{d}}
\newcommand{\efrak}{\mathfrak{e}}
\newcommand{\ffrak}{\mathfrak{f}}
\newcommand{\gfrak}{\mathfrak{g}}
\newcommand{\hfrak}{\mathfrak{h}}
\newcommand{\ifrak}{\mathfrak{i}}
\newcommand{\jfrak}{\mathfrak{j}}
\newcommand{\kfrak}{\mathfrak{k}}
\newcommand{\lfrak}{\mathfrak{l}}
\newcommand{\mfrak}{\mathfrak{m}}
\newcommand{\nfrak}{\mathfrak{n}}
\newcommand{\ofrak}{\mathfrak{o}}
\newcommand{\pfrak}{\mathfrak{p}}
\newcommand{\qfrak}{\mathfrak{q}}
\newcommand{\rfrak}{\mathfrak{r}}
\newcommand{\sfrak}{\mathfrak{s}}
\newcommand{\tfrak}{\mathfrak{t}}
\newcommand{\ufrak}{\mathfrak{u}}
\newcommand{\vfrak}{\mathfrak{v}}
\newcommand{\wfrak}{\mathfrak{w}}
\newcommand{\xfrak}{\mathfrak{x}}
\newcommand{\yfrak}{\mathfrak{y}}
\newcommand{\zfrak}{\mathfrak{z}}

\newcommand{\gab}{g_{\alpha\beta}}
\newcommand{\gabu}{g^{\alpha\beta}}
\newcommand{\Bab}{B_{\alpha\beta}}
\newcommand{\Babu}{B^{\alpha\beta}}
\newcommand{\Xa}{X^\alpha}
\newcommand{\Xb}{X^\beta}

\newcommand{\al}{\alpha}

\newcommand{\bet}{\beta}
\newcommand{\G}{\Gamma}
\newcommand{\g}{\gamma}
\newcommand{\ga}{\gamma}
\newcommand{\de}{\delta}
\newcommand{\De}{\Delta}
\newcommand{\m}{\mu}
\newcommand{\n}{\nu}
\newcommand{\la}{\lambda}
\newcommand{\La}{\Lambda}
\newcommand{\s}{\sigma}
\newcommand{\si}{\sigma}
\newcommand{\Si}{\Sigma}
\newcommand{\tht}{\theta}
\newcommand{\Tht}{\theta}
\newcommand{\om}{\omega}
\newcommand{\Om}{\Omega}
\newcommand{\ep}{\epsilon}
\newcommand{\vep}{\varepsilon}
\newcommand{\ve}{\varepsilon}
\newcommand{\ka}{\kappa}
\newcommand{\io}{\iota}
\newcommand{\ze}{\zeta}

\newcommand{\pd}{\partial}
\newcommand{\parens}[1]{\left(#1\right)}
\newcommand{\bracke}[1]{\left[#1\right]}
\newcommand{\pr}[1]{\left(#1\right)}
\newcommand{\br}[1]{\left[#1\right]}
\newcommand{\cbr}[1]{\left\{#1\right\}}
\newcommand{\crb}[1]{\left\{#1\right\}}
\newcommand{\norm}[1]{\left\Vert#1\right\Vert}
\newcommand{\detm}[1]{\left|#1\right|}
\newcommand{\abs}[1]{\left\vert#1\right\vert}
\newcommand{\set}[1]{\left\{#1\right\}}
\newcommand{\eps}{\varepsilon}
\newcommand{\To}{\longrightarrow}
\newcommand{\OP}[1]{\mathcal{#1}}
\newcommand{\derv}[2]{\frac{d#1}{d#2}}
\newcommand{\nderv}[3]{\frac{d^{#1}#2}{{d#3}^#1}}
\newcommand{\psig}[1]{\sigma_#1}
\newcommand{\thet}{\theta}
\newcommand{\Grf}{\widehat{\G}}
\newcommand{\Qd}[1]{Q_{(#1)}}
\newcommand{\RC}[2]{{\OP{R}^{#1}}_{#2}}
\newcommand{\RB}[2]{\overline{\OP{R}}_{#1}^{\phantom{#1}{#2}}}
\newcommand{\RTB}[2]{\OP{R}_{\bar{#1}}^{\phantom{#1}{\bar{#2}}}}
\newcommand{\RBE}[2]{e^{-i \thet_{(#1 #2)}} {r_{#1}}^{#2}}
\newcommand{\ket}{\rangle}
\newcommand{\bra}{\langle}
\newcommand{\bipd}{\stackrel{\leftrightarrow}{\partial}}
\newcommand{\pbr}[1]{\left\{#1\right\}_{\textrm{PB}}}
\newcommand{\dbr}[1]{\left\{#1\right\}_{\textrm{DB}}}
\newcommand{\deloQ}{\delta^0_Q}
\newcommand{\delIQ}{\delta^1_Q}
\newcommand{\Diff}{\textrm{Diff}}
\newcommand{\Map}{\textrm{Map}}
\newcommand{\subst}[2]{\cbr{#1 \atop #2}}

\newcommand{\Tr}{\textrm{Tr}}
\newcommand{\tr}{\textrm{tr}}
\newcommand{\str}{\textrm{str}}
\newcommand{\STr}{\textrm{STr}}
\newcommand{\Det}{\textrm{Det}}
\newcommand{\Id}{\mathbb{I}}

\newcommand{\del}{\nabla}

\newcommand{\upket}{|\!\!\uparrow\rangle}
\newcommand{\dnket}{|\!\!\downarrow\rangle}
\newcommand{\upbra}{\langle\uparrow|}
\newcommand{\dnbra}{\langle\downarrow|}
\newcommand{\expect}[1]{\langle #1 \rangle}
\newcommand{\Psiket}{|\Psi\rangle}
\newcommand{\Psizerket}{|\Psi^{(0)}\rangle}
\newcommand{\Psioneket}{|\Psi^{(1)}\rangle}
\newcommand{\Psibra}{\langle\Psi|}
\newcommand{\Psizerbra}{\langle\Psi^{(0)}|}
\newcommand{\Psionebra}{\langle\Psi^{(1)}|}
\newcommand{\Psitwobra}{\langle\Psi^{(2)}|}
\newcommand{\Psitwoket}{|\Psi^{(2)}\rangle}

\newcommand{\epsbar}{\bar{\varepsilon}}
\newcommand{\phibar}{\bar{\phi}}
\newcommand{\chibar}{\bar{\chi}}
\newcommand{\psibar}{\bar{\psi}}
\newcommand{\Psibar}{\bar{\Psi}}
\newcommand{\thtbar}{\bar{\theta}}

\newcommand{\zbar}{\bar{z}}

\newcommand{\Pbar}{\bar{P}}
\newcommand{\Qbar}{\bar{Q}}
\newcommand{\Rbar}{\bar{R}}
\newcommand{\Sbar}{\bar{S}}
\newcommand{\Tbar}{\bar{T}}
\newcommand{\Ubar}{\bar{U}}
\newcommand{\Vbar}{\bar{V}}
\newcommand{\Wbar}{\bar{W}}

\newcommand{\pdbar}{\bar{\partial}}

\newcommand{\Asc}{\mathscr{A}}
\newcommand{\asc}{\mathscr{a}}
\newcommand{\Bsc}{\mathscr{B}}
\newcommand{\Csc}{\mathscr{C}}
\newcommand{\Dsc}{\mathscr{D}}
\newcommand{\Esc}{\mathscr{E}}
\newcommand{\Hsc}{\mathscr{H}}
\newcommand{\Lsc}{\mathscr{L}}
\newcommand{\Msc}{\mathscr{M}}
\newcommand{\Nsc}{\mathscr{N}}
\newcommand{\Osc}{\mathscr{O}}
\newcommand{\Psc}{\mathscr{P}}
\newcommand{\Qsc}{\mathscr{Q}}
\newcommand{\Rsc}{\mathscr{R}}
\newcommand{\Ssc}{\mathscr{S}}
\newcommand{\Tsc}{\mathscr{T}}
\newcommand{\Usc}{\mathscr{U}}
\newcommand{\usc}{\mathscr{u}}
\newcommand{\Vsc}{\mathscr{V}}
\newcommand{\Wsc}{\mathscr{W}}

\newcommand{\Acl}{\mathcal{A}}
\newcommand{\Bcl}{\mathcal{B}}
\newcommand{\Ccl}{\mathcal{C}}
\newcommand{\Dcl}{\mathcal{D}}
\newcommand{\Ecl}{\mathcal{E}}
\newcommand{\Hcl}{\mathcal{H}}
\newcommand{\Mcl}{\mathcal{M}}
\newcommand{\Ncl}{\mathcal{N}}
\newcommand{\Ocl}{\mathcal{O}}
\newcommand{\Pcl}{\mathcal{P}}
\newcommand{\Qcl}{\mathcal{Q}}
\newcommand{\Rcl}{\mathcal{R}}

\newcommand{\eb}{\mathbf{e}}
\newcommand{\gb}{\mathbf{g}}
\newcommand{\hb}{\mathbf{h}}
\newcommand{\kb}{\mathbf{k}}

\newcommand{\ahat}{\hat{a}}
\newcommand{\ghat}{\hat{g}}
\newcommand{\nhat}{\hat{n}}
\newcommand{\phat}{\hat{p}}
\newcommand{\uhat}{\hat{u}}
\newcommand{\vhat}{\hat{v}}
\newcommand{\xhat}{\hat{x}}
\newcommand{\yhat}{\hat{y}}
\newcommand{\zhat}{\hat{z}}

\newcommand{\Ahat}{\widehat{A}}
\newcommand{\Bhat}{\hat{B}}
\newcommand{\Dhat}{\widehat{D}}
\newcommand{\Chat}{\widehat{C}}
\newcommand{\Ehat}{\widehat{E}}
\newcommand{\Fhat}{\widehat{F}}
\newcommand{\Ghat}{\widehat{G}}
\newcommand{\Hhat}{\widehat{H}}
\newcommand{\Ihat}{\widehat{I}}
\newcommand{\Jhat}{\widehat{J}}
\newcommand{\Khat}{\widehat{K}}
\newcommand{\Lhat}{\widehat{L}}
\newcommand{\Mhat}{\widehat{M}}
\newcommand{\Nhat}{\widehat{N}}
\newcommand{\Ohat}{\widehat{O}}
\newcommand{\Phat}{\widehat{P}}
\newcommand{\Qhat}{\widehat{Q}}
\newcommand{\Rhat}{\widehat{R}}
\newcommand{\Shat}{\widehat{S}}
\newcommand{\That}{\widehat{T}}
\newcommand{\Uhat}{\widehat{U}}
\newcommand{\Vhat}{\widehat{V}}
\newcommand{\What}{\widehat{W}}
\newcommand{\Xhat}{\widehat{X}}
\newcommand{\Yhat}{\widehat{Y}}
\newcommand{\Zhat}{\widehat{Z}}

\newcommand{\gahat}{\hat{\gamma}}
\newcommand{\sighat}{\hat{\sigma}}
\newcommand{\omhat}{\hat{\omega}}
\newcommand{\Omhat}{\hat{\Omega}}
\newcommand{\phihat}{\hat{\phi}}
\newcommand{\Phihat}{\hat{\Phi}}
\newcommand{\pihat}{\hat{\pi}}
\newcommand{\Pihat}{\hat{\Pi}}
\newcommand{\xihat}{\hat{\xi}}
\newcommand{\Xihat}{\hat{\Xi}}

\newcommand{\iqhat}{\hat{\mathbf{i}}}
\newcommand{\jqhat}{\hat{\mathbf{j}}}
\newcommand{\kqhat}{\hat{\mathbf{k}}}

\newcommand{\Atilde}{\widetilde{A}}
\newcommand{\Btilde}{\widetilde{B}}
\newcommand{\Ctilde}{\widetilde{C}}
\newcommand{\Dtilde}{\widetilde{D}}
\newcommand{\Ktilde}{\widetilde{K}}
\newcommand{\Ltilde}{\widetilde{L}}
\newcommand{\Mtilde}{\widetilde{M}}
\newcommand{\Ntilde}{\widetilde{N}}
\newcommand{\Otilde}{\widetilde{O}}
\newcommand{\Ptilde}{\widetilde{P}}
\newcommand{\Qtilde}{\widetilde{Q}}
\newcommand{\Rtilde}{\widetilde{R}}
\newcommand{\Stilde}{\widetilde{S}}

\newcommand{\atilde}{\tilde{a}}
\newcommand{\btilde}{\tilde{B}}
\newcommand{\ctilde}{\tilde{c}}
\newcommand{\dtilde}{\tilde{d}}
\newcommand{\etilde}{\tilde{e}}
\newcommand{\ftilde}{\tilde{f}}
\newcommand{\ktilde}{\tilde{k}}
\newcommand{\ltilde}{\tilde{l}}
\newcommand{\mtilde}{\tilde{m}}
\newcommand{\ntilde}{\tilde{n}}
\newcommand{\otilde}{\tilde{o}}
\newcommand{\ptilde}{\tilde{p}}
\newcommand{\qtilde}{\tilde{q}}
\newcommand{\rtilde}{\tilde{r}}
\newcommand{\stilde}{\tilde{s}}

\newcommand{\gatild}{\tilde{\gamma}}
\newcommand{\psitilde}{\tilde{\psi}}

\newcommand{\Acheck}{\check{A}}
\newcommand{\Bcheck}{\check{B}}
\newcommand{\Ccheck}{\check{C}}
\newcommand{\Dcheck}{\check{D}}
\newcommand{\Echeck}{\check{E}}
\newcommand{\Fcheck}{\check{F}}
\newcommand{\Gcheck}{\check{G}}
\newcommand{\Hcheck}{\check{H}}
\newcommand{\Icheck}{\check{I}}
\newcommand{\Jcheck}{\check{J}}
\newcommand{\Kcheck}{\check{K}}
\newcommand{\Lcheck}{\check{L}}
\newcommand{\Mcheck}{\check{M}}
\newcommand{\Ncheck}{\check{N}}
\newcommand{\Ocheck}{\check{O}}
\newcommand{\Pcheck}{\check{P}}
\newcommand{\Qcheck}{\check{Q}}
\newcommand{\Rcheck}{\check{R}}
\newcommand{\Scheck}{\check{S}}

\newcommand{\xdot}{\dot{x}}

\newcommand{\Xdot}{\dot{X}}

\newcommand{\xidot}{\dot{\xi}}

\newcommand{\Svec}{\vec{S}}

\newcommand{\sigvec}{\vec{\sigma}}

\newcommand{\nab}{\nabla}

\newcommand{\eff}{\textrm{eff}}

\newcommand{\Ord}{\mathscr{O}}

\newcommand{\overrt}[1]{\frac{1}{\sqrt{#1}}}

\newtheorem{point}{Point}

\newcommand{\End}{{\mathrm{End}}}

%----- Begin Main Text -------

\begin{flushright}
\break
math-ph/0708.3687\\
DAMTP-2007-78\\
DCPT-07/47\\[3mm]
\end{flushright}
\vspace{1cm}
\begin{center}
{\Large {\bf Multiplicity in Supersymmetric Spin Chains}}

\vspace{0.8cm} {\large David Kagan$^a$
                       and Charles A. S. Young$^b$}\\[3mm]
{\em $^a$DAMTP, Centre for Mathematical Sciences, University of Cambridge,\\
Wilberforce Road, Cambridge CB3 0WA, UK}\\[3mm]
{ \emph{$^b$Department of Mathematical Sciences, University of Durham\\
South Road, Durham DH1 3LE, UK}}\\
{\small E-mail: {\tt d.kagan@damtp.cam.ac.uk, charles.young@durham.ac.uk}}
\end{center}

\vskip 0.15in

\begin{abstract}
We discuss a simple procedure for obtaining new integrable spin chains from old by replacing each single state of the original model by some collection of states. This works whenever the Lax matrix of the chain has a certain form. The simplest example is the $\su(n)$ XX model. We apply the techniques of the nested algebraic Bethe ansatz to solve such systems, in the bosonic and supersymmetric cases.

\end{abstract}

\section{Introduction}

In \cite{suXXModels}, Maassarani and Mathieu introduced generalizations of the usual two-state XX spin chain, motivatated in part by the desire to construct versions of the Hubbard model with $\su(n)$ symmetry \cite{SUNHubbard,LaxSUNHub,sun2,MartinsSUNHub}.
Following \cite{suXXModels} these $\su(n)$ XX models were further generalized in  \cite{XXC,AmModels} and it emerged that the underlying idea is rather simple. For a certain class of $R$-matrices it is possible to introduce ``multiplicity''. This means, roughly speaking, replacing each state of a given spin chain by a collection of states, all equivalent from the point of view of the spin chain interactions --- we will make this more precise in section 2. In \cite{AmModels}, this procedure was applied specifically to the $R$-matrix of $A_m$ in the fundamental representation, but it is not limited to it.

The main goal of the present work is to carry out a similar construction in the case of supersymmetric $R$-matrices and spin chains. We first review, in section 2, the introduction of multiplicity into XXZ-like integrable systems, focusing on identifying the features needed to make the construction work. Lax operators, monodromy, and transfer matrices are defined. In section 3 we examine the algebraic Bethe ansatz for the new models.

In section 4 we turn to the supersymmetric case and add multiplicity to integrable models associated with the $\su(m|\,n)$ superalgebras. The form of the algebraic Bethe ansatz for these models is similar to those of the usual $\su(m|\,n)$ chain \cite{SusyABA}, but with modifications due to the additional degrees of freedom. It will turn out that this alters the spectrum of the monodromy matrix and leads to a degeneracy of eigenstates just as in the purely bosonic case.

\section{$A_{m-1}$-type Integrable Models With Multiplicity}

As a warm-up, and to establish notation, we shall consider first bosonic models similar to those in
\cite{AmModels}, before we turn to the supersymmetric case in section \ref{susy}. Let
$\{E_I:I=0,\dots, m-1\}$ be the standard basis of $V=\mathbb C^m$, $\{\Omega^I\}$  the dual basis of
$V^*$, and $E_I{}^J = E_I\otimes \Omega^J\in V \otimes V^*\cong \End(V)$. Suppose $R^o(u)\in
\End(V\otimes V)$ is an $R$-matrix obeying the usual Yang-Baxter equation with spectral
parameter\footnote{We follow the standard notation of \cite{FRT}, so that for example $R_{13}(u)$
denotes the element of $\End(V)^{\otimes 3}$ \be \sum_{I,J,K,L=0}^{m-1} R^I{}_J{}^K{}_L(u) E_I{}^J
\otimes 1 \otimes E_K{}^L .\ee}
\begin{equation}
R^o_{12}(u-v) R^o_{13}(u-w) R^o_{23}(v-w) = R^o_{23}(v-w) R^o_{13}(u-w) R^o_{12}(u-v),
\end{equation}
which, moreover, is of the particular form 
\begin{equation}
R^o(\la) = \sum_{I,J=0}^{m-1} \pr{R^{J}{}_{I}{}^I{}_{J}(\la)\,{E_J}^I \otimes {E_I}^J +
(1-\delta_{IJ})R^I{}_I{}^J{}_J(\la)\,{E_I}^I \otimes {E_J}^J}\label{Ro}.
\end{equation}
That is, in each non-zero element of the $R$-matrix, the set of `in' indices is the same as the set
of `out' indices. The usual trigonometic $R$-matrix of $A_{m-1}$ \cite{Jimbo} is of this type, and was the particular case considered in \cite{AmModels}. From the point of view of statistical mechanics this property originates in the ice rule obeyed by the Boltzmann weights.

Given any such $R$-matrix, the ``multiplicity'' models of \cite{AmModels} are defined by replacing each individual basis state by a collection of states. Thus, for every $E_I$, introduce new basis states $\{e_{a_I} : a_I \in \Asc_I\}$, where $\Asc_I$ is an index set of cardinality $n_I\in \mathbb N$. Let $\{\omega^{a_I}:a_I\in\Asc_I\}$ be the dual basis, write $e_{a_I}{}^{b_J} = e_{a_I}\otimes \omega^{b_J}$, and then define
\begin{equation}
R(\la) = \sum_{I,J=0}^{m-1} \mathop{\sum_{a_I\in \Asc_I}}_{b_J \in \Asc_J}
\pr{R^J{}_I{}^I{}_J(\la)\,{e_{b_J}}^{a_I} \otimes {e_{a_I}}^{b_J} +
(1-\delta_{IJ})R^I{}_I{}^J{}_J(\la)\,{e_{a_I}}^{a_I} \otimes {e_{b_J}}^{b_J}}.
\end{equation}
One way to see that this new $R$-matrix also obeys the Yang-Baxter equation is to note that
$R^o(\lambda)$ may be written purely in terms of the combination $E_I \otimes \Omega^I$ (no sum on
$I$), with the tensor factors inserted in appropriate slots. The new $R$-matrix is obtained by
performing the replacement
\be E_I \otimes \Omega^I \quad \longrightarrow \quad 
    \sum_{a_I\in \Asc_I} e_{a_I} \otimes \omega^{a_I},\ee
so it suffices to observe that these objects obey the same rule under internal contraction: on the one hand
\be  E_I \,\, \left< \Omega^I , E_J \right> \otimes \Omega^J = \delta_{IJ} E_I \otimes \Omega^I,\ee
and on the other, since the index sets $\Asc_I$ are disjoint,
\be \sum_{a_I\in \Asc_I}  e_{a_I} \,\, \left< \omega^{a_I}, \sum_{b_J \in \Asc_J} 
     e_{b_J} \right>  \otimes\omega^{b_J}
 =\sum_{a_I\in \Asc_I}  \sum_{b_J \in \Asc_J} \delta_{a_I b_J} e_{a_I} \otimes \omega^{b_J}
 =\delta_{IJ} \sum_{a_I\in \Asc_I} e_{a_I} \otimes \omega^{a_I}.\ee

We consider a spin chain of length $p_0$, and take as our Lax operator (see e.g. \cite{Kundu} for a
review)
\begin{equation}
L_{\al x}(\mu) = R_{\al x}(\mu - \la^0_x),
\end{equation}
where $x\in\{1,2,\dots,p_0\}$ denotes the space of the $x$th particle on the chain, while the index
$\al$ denotes the auxiliary space. The $\la^0_x$ appearing in the arguments of the $R$-matrices
parametrize the inhomogeneities at each position of the chain. From the Lax operator, we construct
the monodromy matrix
\begin{equation}
T_\al(\mu) = L_{\al p_0}(\mu) L_{\al (p_0-1)}(\mu) \cdots L_{\al 1}(\mu).
\end{equation}
The transfer matrix is the partial trace of the monodromy matrix: $\tau(\mu) = \tr_\al\,T_\al(\mu)$.
Notice that the real difference between the model with multiplicity and the usual model the lies
here: the monodromy matrix makes explicit reference to the indices $a_I$, and the notational trick
used above to recast the new $R$-matrix in the same form as the original one cannot be used here to
simply reduce the new model to the old one.

\section{Algebraic Bethe Ansatz}
The algebraic Bethe ansatz for these models has the same structure as the one developed by
Maassarani for his $A_m$ models in \cite{AmModels}. In fact, the models of \cite{AmModels} are
subsumed by the more general framework here, though we have not included the parameters
Maassarani denotes as $x_{\al_i\beta_j}$, which in our notation would be $x_{a_I b_J}$. Adding these
parameters is not difficult, but we will work with them set equal to unity.

To proceed with the algebraic Bethe ansatz, choose a pseudo-vaccum:
\begin{equation}
|\Omega_0\ket = |\ahat_0 \cdots \ahat_0\ket,
\end{equation}
where $\ahat_0 \in \Asc_0$ and the capital Latin indices now split into $I = (0,i)$. Let $a_I(\la) =
R_{II}^{II}(\la)$, $b_{IJ}(\la) = R_{IJ}^{IJ}(\la)$ and $c_{IJ}(\la) = R_{IJ}^{JI}(\la)$ for $I\neq
J$. The choice of pseudo-vacuum singles out of various elements of the monodromy matrix. Let $A(\la)
= T_{\ahat_0}^{\ahat_0}(\la)$ and $B_{a_J}(\la) = T_{b_J}^{\ahat_0}(\la)$. It is not hard to check
that on the pseudo-vaccum we have
\begin{eqnarray}
A(\mu)|\Omega_0\ket &=& \prod_{x=1}^{p_0} a_0(\mu - \la^0_x) |\Omega_0\ket,\nn\\
T_{a_i}^{a_i}|\Omega_0\ket &=& \prod_{x=1}^{p_0} b_{i0}(\mu - \la^0_x) |\Omega_0\ket.
\end{eqnarray}
The only other operators that do not annihilate the state are
\begin{eqnarray}
B_{a_0}(\la)|\Omega_0\ket &=& \prod_{x=1}^{p_0} a_0(\mu - \la^0_x) |a_0 \ahat_0 \cdots
\ahat_0\ket,\nn\\
B_{a_i}(\la)|\Omega_0\ket &=& \sum_{x=1}^{p_0}
\pr{\prod_{z=x+1}^{p_0}a_0(\mu-\la^0_x)}c_{i0}(\mu-\la^0_x)\pr{\prod_{y=1}^{x-1} b_{i0}(\mu -
\la^0_y)} |\ahat_0 \cdots \overbrace{a_i}^x \cdots \ahat_0\ket.\nn\\
\end{eqnarray}
The second of the above two relations is what we will use to construct the other eigenstates of the
transfer matrix. These states will be of the form
\begin{equation}
|\Psi_1\ket = \sum_{i=1}^{m-1} \sum_{c_{k_1} \in \Asc_{k_1},\ldots,c_{k_{p_1}} \in \Asc_{k_{p_1}}}
F^{c_{k_1} c_{k_2} \cdots c_{k_{p_1}}} B_{c_{k_1}}(\la_1^1) B_{c_{k_2}}(\la_2^1) \cdots
B_{c_{k_{p_1}}}(\la_{p_1}^1)|\Omega_0\ket,
\end{equation}
where $F$ is to be determined. We break the analysis of the action of the transfer matrix up into
$m-1$ levels and at each level aside from the final one, the procedure remains essentially the same.
First, work out the action of $T_{a_0}^{a_0}(\mu)$ on $|\Psi_1\ket$, where $a_0 \neq \ahat_0$. These
elements of the monodromy matrix annihilate $|\Omega_0\ket$, but it turns out that they preserve
states of the form $|a_{i_1} \cdots a_{i_{p_1}}\ket$ when $p_1 = p_0$. This is an effect of the
multiplicity that does not occur in the usual models. Explicitly we have
\begin{eqnarray*}
T_{a_0}^{a_0}(\mu)|a_{i_1}\cdots a_{i_{p_0}}\ket 
&=& \sum_{k,j=1}^{m-1}
\mathop{\sum_{b_{j_1}\in\Asc_{j_1},\ldots,b_{j_{p_0}}\in\Asc_{j_{p_0}}}}_{c_{k_2}\in\Asc_{k_2},\ldots,c_{k_{p_0}}\in\Asc_{c_{p_0}}}
R^{a_0\ \ b_{j_{p_0}}}_{c_{k_{p_0}} a_{i_{p_0}}}(\mu - \la_{p_0}^0) \cdots R^{c_{i_2} b_{j_1}}_{a_0
a_{i_1}}(\mu - \la_1^0)|b_{j_1} \cdots b_{j_{p_0}}\ket \\
&=& \prod_{x=1}^{p_0} b_{0i}(\mu - \la_x^0)|a_{i_1} \cdots a_{i_{p_0}}\ket.
\end{eqnarray*}
As a result, the action of $T_{a_0}^{a_0}$ on $|\Psi_1\ket$ is
\begin{equation}
T_{a_0}^{a_0}(\mu)|\Psi_1\ket = \delta_{p_0 p_1}\prod_{x=1}^{p_1} b_{0i}(\mu - \la_x^0)|\Psi_1\ket.
\end{equation}
The action of remaining components of monodromy matrix can be dealt with using the $RTT$-equations,
\be R_{\al\beta}(\la-\mu) T_\al(\la) T_\beta(\mu)
   = T_\beta(\mu) T_\al(\la) R_{\al\beta}(\la-\mu),\ee  which, with the matrix indices displayed
explicitly, read
\begin{equation}
\sum_{M_1,M_2} \mathop{\sum_{e_{M_1} \in \Asc_{M_1}}}_{e_{M_2}\in \Asc_{M_2}}
R^{a_{I_1} a_{I_2}}_{e_{M_1} e_{M_2}}(\la - \mu)T_{b_{J_1}}^{e_{M_1}}(\la)T_{b_{J_2}}^{e_{M_2}}(\mu)
=
\sum_{M_1,M_2} \mathop{\sum_{e_{M_1} \in \Asc_{M_1}}}_{e_{M_2}\in \Asc_{M_2}}
T_{e_{M_2}}^{a_{I_2}}(\mu)T_{e_{M_1}}^{a_{I_1}}(\la)R_{b_{J_1} b_{J_2}}^{e_{M_1} e_{M_2}}(\la -
\mu).
\end{equation}
By specifying the various free indices, we may write down the particular relations which interest
us:
\begin{eqnarray}
A(\mu)B_{a_i}(\la) &=& {a_0(\la-\mu)\over b_{i0}(\la-\mu)}B_{a_i}(\la) A(\mu) -
{c_{i0}(\la-\mu)\over b_{i0}(\la-\mu)}B_{a_i}(\mu) A(\la) \\
T_{a_i}^{b_j}(\mu) B_{c_k}(\la) &=& \sum_{r,s=1}^{m-1} \mathop{\sum_{e_r\in\Asc_r}}_{f_t\in\Asc_t}
{R^{\,e_r f_t}_{\,a_i\ c_k}(\mu-\la)\over b_{j0}(\mu-\la)}B_{f_t}(\la) T_{e_r}^{b_j}(\mu)
-{c_{0j}(\mu-\la)\over b_{j0}(\mu-\la)}B_{a_i}(\mu) T_{c_k}^{b_j}(\la).
\end{eqnarray}
Ignoring contributions from the second term on the right-hand-side of either equation, we proceed by
computing first
\begin{equation}
A(\mu)|\Psi_1\ket = 
\prod_{x=1}^{p_0} a_0(\mu - \la^0_x)
\prod_{y=1}^{p_1} {a_0(\la_y^1-\mu)\over b_{0}(\la_y^1 - \mu)} |\Psi_1\ket,
\end{equation}
where we are forced to make the simplification $b_{k0} = b_0$ in order for $|\Psi_1\ket$ to be an
eigenstate of $A(\mu)$. Now compute the action of $T_{a_i}^{a_i}$ on $|\Psi_1\ket$
\begin{eqnarray}
T_{a_i}^{a_i}(\mu)|\Psi_1\ket &=& \prod_{x=1}^{p_0} b_0(\mu-\la_x^0) \prod_{y=1}^{p_1} {1\over
b_0(\mu-\la_y^1)}
\mathop{\sum_{r_1,\ldots,r_{p_1}=1}^{m-1}}_{t_1,\ldots,t_{p_1}=1} 
\mathop{\sum_{e_{r_1}\in\Asc_{r_1},\ldots,e_{r_{p_1}}\in
\Asc_{r_{p_1}}}}_{f_{t_1}\in\Asc_{t_1},\ldots,f_{t_{p_1}}\in \Asc_{t_{p_1}}} \nn\\
&\times& B_{f_{t_1}}(\la_1^1)\cdots B_{f_{t_{p_1}}}(\la_{p_1}^1)|\Omega_0\ket\,
(T^1)^{a_i,f_{t_1}\cdots f_{t_{p_1}}}_{a_i,c_{k_1}\cdots c_{k_{p_1}}}(\mu) F^{c_{k_1} \cdots
c_{k_{p_1}}},
\end{eqnarray}
where the level 1 monodromy matrix is defined as
\begin{equation}
T^1_{\al_1}(\mu) = L_{\al_1 p_1}(\mu-\la_{p_1}^1) \cdots L_{\al_1 1}(\mu-\la_1^1),
\end{equation}
and $\al_1$ refers to the level 1 auxiliary space whose components only run over the indices in the
sets $\cbr{\Asc_i}_{i=1}^{m-1}$. Clearly, we have now come around to an analogous problem one level
up. The undetermined coefficients $F$ appearing in $|\Psi_1\ket$ are in fact, states of a level 1
spin chain acted on by the level 1 monodromy matrix. We wish to choose these to be eigenstates of
the diagonal blocks of the level 1 monodromy matrix:
\begin{equation}
(T^1)_{a_i}^{a_i}(\mu)F = (\La^1)_{a_i}^{a_i}(\mu) F,
\end{equation}
where $(\La^1)_{a_i}^{a_i}(\mu)$ is the level 1 eigenvalue. Given this, the expression for the
action of $T_{a_i}^{a_i}|\Psi_1\ket$ simplifies to
\begin{equation}
T_{a_i}^{a_i}(\mu)|\Psi_1\ket = \prod_{x=1}^{p_0} b_0(\mu-\la_x^0) \prod_{y=1}^{p_1} {1\over
b_0(\mu-\la_y^1)} (\La^1)_{a_i}^{a_i}(\mu)|\Psi_1\ket.
\end{equation}

So we have deduced the eigenvalue up to the next level in the analysis. It is
\begin{eqnarray*}
\La(\mu) &=& \delta_{p_0 p_1} (n_0 - 1) \prod_{x=1}^{p_0} b_0(\mu-\la_x^0) 
          + \prod_{x=1}^{p_0} a_0(\mu - \la^0_x)\prod_{y=1}^{p_1} {a_0(\la_y^1-\mu)\over
b_{0}(\la_y^1 - \mu)}\\
         &+& \prod_{x=1}^{p_0} b_0(\mu-\la_x^0) \prod_{y=1}^{p_1} {1\over b_0(\mu-\la_y^1)}
\La^1(\mu),
\end{eqnarray*}
where $\La^1 = \sum_i \sum_{a_i \in \Asc_i} (\La^1)_{a_i}^{a_i}$. The factor of $n_0 - 1$ appearing
in the first term arises from the part of the trace that sums over all the states in $\Asc_0$ except
for $\ahat_0$. The higher levels have the same structure, so the higher level eigenvalues are
\begin{eqnarray}
\La^k(\mu) &=& \delta_{p_k p_{k+1}} (n_k - 1) \prod_{x=1}^{p_k} b_k(\mu-\la_x^k) 
          + \prod_{x=1}^{p_k} a_k(\mu - \la^k_x)\prod_{y=1}^{p_{k+1}} {a_k(\la_y^{k+1}-\mu)\over
b_k(\la_y^{k+1} - \mu)}\nn\\
         &+& \prod_{x=1}^{p_k} b_k(\mu-\la_x^k) \prod_{y=1}^{p_{k+1}} {1\over b_k(\mu-\la_y^{k+1})}
\La^{k+1}(\mu).
\end{eqnarray}
Here $k$ runs over $0,\ldots,m-2$. We have also defined $b_k(\mu) = b_{ki}(\mu)$ for all
$i=1,\ldots,m-1$. Also note that throughout we have left implicit the dependencies on the
inhomogeneities $\cbr{\la_1^k,\ldots,\la_{p_k}^k}$ that appear at each level $k=0,\ldots,m-2$. The
final eigenvalue $\La^{m-1}$ appears when $k=m-2$. Examining the $RTT$-equations for this final
level:
\begin{eqnarray*}
(T^{m-2})_{a}^{b}(\mu) (B^{m-2})_{c}(\la) 
&=& \mathop{\sum_{e\in\Asc_{m-1}}}_{f\in\Asc_{m-1}} {R^{\,ef}_{ac}(\mu-\la)\over
b_{m-1}(\mu-\la)}(B^{m-2})_{f}(\la) (T^{m-2})_{e}^{b}(\mu) \\
&-& {c_{m-1}(\mu-\la)\over b_{m-1}(\mu-\la)}(B^{m-2})_{a}(\mu) (T^{m-2})_{c}^{b}(\la),
\end{eqnarray*}
where all the indices above take values in $\Asc_{m-1}$. The level $m-2$ creation operator is given
by $(B^{m-2})_c = (T^{m-2})^{\ahat_{m-2}}_c$, where $\ahat_{m-2} \in \Asc_{m-2}$ is the chosen level
$m-2$ pseudo-vacuum state. It is not hard to see that for $e,f,a,c\in \Asc_{m-1}$ we have
\begin{equation}
R_{ac}^{\,ef}(\la) = a_{m-1}(\la) \delta_a^f \delta_c^e = a_{m-1}(\la) P_{ac}^{ef},
\end{equation}
where $P$ is the permutation operator. So the equation simplifies to
\begin{eqnarray}
(T^{m-2})_{a}^{b}(\mu) (B^{m-2})_{c}(\la) 
&=& \mathop{\sum_{e\in\Asc_{m-1}}}_{f\in\Asc_{m-1}} {a_{m-1}(\mu-\la)\over
b_{m-1}(\mu-\la)}(B^{m-2})_{f}(\la) (T^{m-2})_{e}^{b}(\mu) P_{ac}^{ef} \nn\\
&-& {c_{m-1}(\mu-\la)\over b_{m-1}(\mu-\la)}(B^{m-2})_{a}(\mu) (T^{m-2})_{c}^{b}(\la).
\end{eqnarray}
We can deduce that the level $m-1$ monodromy matrix is built out of permutation operators:
\begin{equation}\label{UnitShift}
(T^{m-1})_\al(\mu) = P_{\al p_{m-1}} \cdots P_{\al 1} \prod_{x=1}^{p_{m-1}} a(\mu-\la^{m-1}_x).
\end{equation}
The product of permutation operators yields the unit-shift operator for a chain of length $p_{m-1}$.
The eigenvalues of this operator are roots of unity, and thus, the eigenvalues of the level $m-1$
transfer matrix are of the form of a root of unity times the product $\prod_{x=1}^{p_{m-1}}
a(\mu-\la^{m-1}_x)$.

The Bethe equations for this chain are derived by requiring that the residues around the various
$\la$-parameters vanish. Thus for $k=0,\ldots,m-2$ we have
\begin{eqnarray*}
\La^{k+1}(\la^{k+1}_z)
&=&
\prod_{x=1}^{p_k} {a_k(\la^{k+1}_z - \la^k_x)\over b_k(\la^{k+1}_z-\la_x^k)} 
\mathop{\prod_{y=1}^{p_{k+1}}}_{y\neq z} {a_k(\la_y^{k+1}-\la^{k+1}_z) b_k(\la^{k+1}_z-\la_y^{k+1})
\over b_k(\la_y^{k+1} - \la^{k+1}_z)},
\end{eqnarray*}
where we have assumed that $b_k(\mu) \to 0$ when $\mu \sim 0$. Substituting in the expression for
$\La^{k+1}(\la^{k+1}_z)$ and rearranging yields
\begin{eqnarray}
\prod_{y=1}^{p_{k+2}} {a_{k+1}(\la_y^{k+2}-\la^{k+1}_z)\over b_{k+1}(\la_y^{k+2} - \la^{k+1}_z)}
\mathop{\prod_{y=1}^{p_{k+1}}}_{y\neq z} {a_{k+1}(\la^{k+1}_z-\la_y^{k+1})
b_k(\la_y^{k+1}-\la^{k+1}_z) 
\over a_k(\la_y^{k+1} - \la^{k+1}_z) b_k(\la^{k+1}_z-\la_y^{k+1})}
\prod_{x=1}^{p_k} {b_k(\la^{k+1}_z - \la^k_x)\over a_k(\la^{k+1}_z-\la_x^k)}
={1\over a_{k+1}(0)}. \nn\\
\end{eqnarray}

\section{$A_{\,m-1|\,n-1}$ Models}\label{susy}

We now turn to our main topic of interest, namely the introduction of multiplicity to supersymmetric chains. Suppose henceforth that $V$ is a $\mathbb Z_2$-graded vector space and that our basis is chosen such that each basis vector $E_I$ has a definite grade $\gr I \in \{0,1\}$. Because vectors of grade 0 (respectively 1) are taken to obey bose (fermi) exchange statistics, the tensor product becomes braided (see for example \cite{Majid,ZhangHopf,ZhangElliptic}) in a mild fashion: 
\be (1\otimes E_I) (E_J \otimes 1) = (-1)^{\gr I \gr J} E_J \otimes E_I .\label{stats}\ee
This braiding means that there is some subtlety when taking tensor products of elements of $\End(V)$, essentially because the isomorphism between $\End(V)\otimes \End(V)$ and $\End(V\otimes V)$ is no longer quite trivial. Let us regard $E_I{}^J = E_I\otimes \Omega^J \in \End(V)$ as acting on $V^*$ from the right.\footnote{This will allow our conventions to match those of \cite{SusyABA}; we could instead consider the action from the left on $V$ but the resulting definition (\ref{SUSYox}) would be rather different.} 
\be E_I{}^J: \Omega^M\longrightarrow \Omega^M E_I{}^J = \delta_M^I \Omega^J.\ee 
We then define $\sox$ by the demand that, acting from the right on $V^*\otimes V^*$, $E_I{}^K \sox E_J{}^L$ send
\be \Omega^M \otimes \Omega^N \longrightarrow  \delta^M_I \delta^N_J \,\Omega^K\otimes \Omega^L.\ee
Noting that $E_I{}^K$ is of grade $\gr I + \gr K\mod 2$, we have
\bea (\Omega^M \otimes \Omega^N) (E_I{}^K\otimes E_J{}^L) 
  &=& (-1)^{ \gr N(\gr I + \gr K)} (\Omega^M E_I{}^K\otimes \Omega^N E_J{}^L)  \\
  &=& (-1)^{ \gr J(\gr I + \gr K)} \delta^M_I \delta^N_J \,\,\Omega^K\otimes \Omega^L.\eea
and therefore
\be E_I{}^K \sox E_J{}^L :=  (-1)^{\gr J(\gr I+\gr K)} E_I{}^K\otimes E_J{}^L.\label{SUSYox}\ee
The definition extends naturally to more copies of $\End(V)$, by including signs as needed to ensure that the action from the right on $V^*\otimes\dots \otimes V^*$ is correct. $\sox$ is often called the ``graded'' or ``supersymmetric'' tensor product, although with the present conventions this is somewhat misleading because the grading is already built into $\otimes$, as in (\ref{stats}).  

Given these preliminaries, it is possible to proceed much as in the bosonic case, being careful to allow for the extra signs in (\ref{SUSYox}). Consider an $R$-matrix obeying the graded Yang-Baxter equation\footnote{The advantage of using $\sox$ appears here: this is an equation in $\End(V\otimes V \otimes V)\cong \End(V) \sox \End(V) \sox \End(V)$, and in component form the multiplication is straightforward matrix multiplication. The extra signs are wrapped up in the definition of e.g. $R_{12} = R\sox 1$.}
\begin{equation}
R^o_{12}(u-v) R^o_{13}(u-w) R^o_{23}(v-w) = R^o_{23}(v-w) R^o_{13}(u-w) R^o_{12}(u-v),
\end{equation}
which is of the particular form
\begin{equation}
R^o(\la) = \sum_{I,J=0}^{m+n-1} \pr{r_{IJ}(\la) {E_I}^I \sox {E_J}^J + (-1)^{\gr I \gr J} (1-\delta_{IJ}) t_{IJ}(\la) {E_I}^J \sox {E_J}^I}.
\end{equation}
One class of examples are the $\su(m|n)$ $R$-matrices of \cite{ZhangHopf,ZhangElliptic}, whose
Boltzmann weights are
\begin{eqnarray}
r_{II}(\la) = a_I(\la) &=& {q^{2(1-\gr I )}-q^{2 \gr I } e^{2\la} \over q^2 - e^{2\la}} \label{afn} \\
r_{IJ}(\la) = b(\la) &=& {q(1-e^{2\la})\over q^2 - e^{2\la}}, \ \ \ I \neq J \label{bfn} \\
t_{IJ}(\la) = c(\la) &=& {(q^2-1)e^{2\la} \over q^2-e^{2\la}}, \ \ \ I > J \label{cfn} \\
t_{IJ}(\la) = d(\la) &=& {(q^2-1) \over q^2-e^{2\la}}, \ \ \ I < J. \label{dfn}
\end{eqnarray}
The algebraic Bethe ansatz for models built from this $R$-matrix (including integrable impurities
involving dual representations) was performed in \cite{SusyABA}. 

Now we introduce multiplicity to the models. Once more, introduce the index sets $\cbr{\Asc_I}_{I=0}^{m+n-1}$, and the vector and dual vector bases $\cbr{e_{a_I}}_{a_I \in \Asc_I\!, I=0}^{m+n-1}$, $\cbr{\omega^{a_I}}_{a_I \in \Asc_I\!, I=0}^{m+n-1}$. These have precisely the same properties as before, and again the modified $R$-matrix
\begin{equation}
R(\la) = \sum_{I,J=0}^{m-1} \mathop{\sum_{a_I\in \Asc_I}}_{b_J \in \Asc_J}
\pr{r_{IJ}(\la)\,{e_{a_I}}^{a_I} \sox {e_{b_J}}^{b_J}  +
(-)^{|I||J|}(1-\delta_{IJ}) t_{IJ}(\la) \,{e_{b_J}}^{a_I} \sox {e_{a_I}}^{b_J}}.
\end{equation}
satisfies the Yang-Baxter equation. We choose the Lax operators
\begin{equation}
L_{\al x}(\mu) = R_{\al x}(\mu-\la^0_x)
\end{equation}
and the monodromy matrix for an inhomogeneous chain of length $p_0$ is then
\begin{equation}
T_{\al}(\mu) = L_{\al p_0}(\mu) L_{\al (p_0-1)}(\mu) \cdots L_{\al 1}(\mu).
\end{equation}
The transfer matrix is obtained by taking the supertrace in the auxiliary space $\al$:
\begin{equation}
\tau(\mu) = \str_{\al} T_\al(\mu) = \sum_I \sum_{a_I \in \Asc_I}(-)^{\gr I } T_{a_I}^{a_I}(\mu).
\end{equation}

It follows from the Yang-Baxter equation that $T(\mu)$ satisfies the $RTT$ relations
\begin{equation}
R_{12}(\la - \mu) T_1(\la) T_2(\mu) = T_2(\mu) T_1(\la) R_{12} (\la - \mu),
\end{equation}
With all indices displayed explicitly, these read (some care is needed with the sign in $T_2(\mu)T_1(\la) = (1\sox T(\mu))(T(\la) \sox 1)$ on the right-hand side here)
\be (-1)^{\gr L(\gr J + \gr M)} R^I{}_J{}^K{}_L(\la-\mu) T^{e_J}{}_{c_M} (\la) T^{f_L}{}_{d_N}(\mu) 
 = (-1)^{\gr J(\gr I+\gr P)} T^{b_J}{}_{h_Q}(\mu) T^{a_I}{}_{g_P}(\la) R^P{}_M{}^Q{}_N(\la-\mu).\ee

The algebraic Bethe ansatz for the supersymmetric XXZ models is very similar to that of the ordinary
XXZ models, the main difference being a matter of including the signs in the equation above. This means that the algebraic Bethe ansatz for $A_{m|n}$ models will be similar to that of the previous purely bosonic models. Just as before we choose a pseudo-vaccuum:
\begin{equation}
|\Omega_0\ket = |\ahat_0 \cdots \ahat_0\ket,
\end{equation}
where $\ahat_0 \in \Asc_0$ and the capital Latin indices now split into $I = (0,i)$. The elements of
the monodromy matrix that this singles out are $A(\la) = T_{\ahat_0}^{\ahat_0}(\la)$ and
$B_{b_J}(\la) = T_{b_J}^{\ahat_0}(\la)$. We are interested in the action of the following operators
on the pseudo-vacuum
\begin{eqnarray}
A(\mu)|\Omega_0\ket &=& \prod_{x=1}^{p_0} a_0(\mu - \la^0_x) |\Omega_0\ket,\nn\\
T_{a_i}^{a_i}|\Omega_0\ket &=& \prod_{x=1}^{p_0} b(\mu - \la^0_x) |\Omega_0\ket, \nn\\
B_{a_i}(\la)|\Omega_0\ket &=& \sum_{x=1}^{p_0} (-)^{\gr i \gr 0}
\pr{\prod_{z=x+1}^{p_0}a_0(\mu-\la^0_z)}c(\mu-\la^0_x)\pr{\prod_{y=1}^{x-1} b(\mu - \la^0_y)}
|\ahat_0 \cdots \overbrace{a_i}^x \cdots \ahat_0\ket.\nn\\
\end{eqnarray}
The other eigenstates of the transfer matrix will be linear combinations of the states
\begin{equation}
|\Psi_1\ket = \sum_{i=1}^{m-1} \sum_{c_{k_1} \in \Asc_{k_1},\ldots,c_{k_{p_1}} \in \Asc_{k_{p_1}}}
F^{c_{k_1} c_{k_2} \cdots c_{k_{p_1}}} B_{c_{k_1}}(\la_1^1) B_{c_{k_2}}(\la_2^1) \cdots
B_{c_{k_{p_1}}}(\la_{p_1}^1)|\Omega_0\ket,
\end{equation}
where $F$ will turn out to be a chain at the next level of nesting. Recall from before that the
multiplicity leads to contributions that do not appear in the standard models. Namely,
\begin{equation}
T_{a_0}^{a_0}(\mu)|\Psi_1\ket = \delta_{p_0 p_1}\prod_{x=1}^{p_1} b(\mu - \la_x^0)|\Psi_1\ket.
\end{equation}
We may read off the relations that we need from the $RTT$ equations
\begin{eqnarray}
A(\mu)B_{a_i}(\la) &=& {a_0(\la-\mu)\over b(\la-\mu)}B_{a_i}(\la) A(\mu) -
(-)^{\gr 0}{c(\la-\mu)\over b(\la-\mu)}B_{a_i}(\mu) A(\la) \\
T_{a_i}^{b_j}(\mu) B_{c_k}(\la) &=& \sum_{r,s=1}^{m-1} \mathop{\sum_{e_r\in\Asc_r}}_{f_t\in\Asc_t}
(-)^{\gr t\pr{\gr r+\gr j }+\gr 0\pr{\gr i+\gr j }}{R^{\,e_r f_t}_{\,a_i\ c_k}(\mu-\la)\over
b(\mu-\la)}B_{f_t}(\la) T_{e_r}^{b_j}(\mu) \nn\\
&-&
(-)^{\gr 0 \gr i+\gr j\pr{\gr 0+\gr i}}{d(\mu-\la)\over b(\mu-\la)}B_{a_i}(\mu) T_{c_k}^{b_j}(\la).
\end{eqnarray}
Observe that the $R$-matrix appearing in the second equation is the one that describes the
$A_{m-1+\gr 0 |n-\gr 0}$ model. Since each additional level (until the final one) is identical in
structure to the initial level, we deduce that the general eigenvalue at each level is
\begin{eqnarray}
\La^k(\mu) &=& (-)^{\gr k}\delta_{p_k p_{k+1}} (n_k - 1) \prod_{x=1}^{p_k} b(\mu-\la_x^k) 
          + (-)^{\gr k}\prod_{x=1}^{p_k} a_k(\mu - \la^k_x)\prod_{y=1}^{p_{k+1}}
{a_k(\la_y^{k+1}-\mu)\over b(\la_y^{k+1} - \mu)}\nn\\
         &+& \prod_{x=1}^{p_k} b(\mu-\la_x^k) \prod_{y=1}^{p_{k+1}} {1\over b(\mu-\la_y^{k+1})}
\La^{k+1}(\mu),
\end{eqnarray}
for $k = 0,\ldots,m+n-2$. The level $k+1$ transfer matrix is defined as
\begin{eqnarray}\label{kTransfer}
\tau^{(k+1)}(\mu)_{c_{s_1} \cdots c_{s_{p_k}}}^{f_{t_1} \cdots f_{t_{p_k}}}
&=& \sum_{i,r_1,\ldots,r_{p_k-1}=k+1}^{m+n-1} \mathop{\sum_{a_i \in \Asc_i}}_{e_{r_1} \in
\Asc_{r_1},\ldots,e_{r_{p_k-1}} \in \Asc_{r_{p_k-1}}}
(-)^{\sum_{x=1}^{p_k-1} \gr 0 \gr r_x + \sum_{x=1}^{p_k}\pr{\gr{s_x} \gr{r_x} + \gr i \gr{s_x}}} \nn\\
& & 
(-)^{\gr 0\gr i(p_k-1)} R^{\,a_i\ \ \ \ f_{t_{p_k}}}_{e_{r_{p_k-1}} c_{s_{p_k}}}(\mu-\la^k_{p_k})
R^{e_{r_{p_k-1}} f_{t_{p_k-1}}}_{e_{r_{p_k-2}} c_{s_{p_k-1}}}(\mu-\la^k_{p_k-1})\cdots
R^{{e_{r_{1}}} f_{t_{1}}}_{a_i\ c_{s_{1}}}(\mu-\la^k_1). \nn\\
\end{eqnarray}
We can derive this transfer matrix from a suitably defined level $k+1$ monodromy matrix
\begin{equation}
T^{k+1}_{\al_{k+1}}(\mu) = L_{\al_{k+1} p_{k+1}}(\mu-\la_{p_{k+1}}^{k+1}) \cdots L_{\al_{k+1}
1}(\mu-\la_1^{k+1}).
\end{equation}
The products in the level $k+1$ auxiliary space whose components only run over the indices in the
sets $\cbr{\Asc_i}_{i=k+1}^{m+n-1}$ and the new supertrace are defined to account for the signs
appearing in (\ref{kTransfer}). The $F$ coefficients at level $k$ are chosen to diagonalize the
level $k+1$ transfer matrix
\begin{equation}
\tau^{k+1}(\mu)F^{k} = \La^{k+1}(\mu) F^k,
\end{equation}
where $\La^{k+1}(\mu)$ is the level $k+1$ eigenvalue.

At the final level the $RTT$-equations are the same as those appearing in the bosonic case
\begin{eqnarray*}
(T^{m-2})_{a}^{b}(\mu) (B^{m-2})_{c}(\la) 
&=& \mathop{\sum_{e\in\Asc_{m-1}}}_{f\in\Asc_{m-1}} {R^{\,ef}_{ac}(\mu-\la)\over
b_{m-1}(\mu-\la)}(B^{m-2})_{f}(\la) (T^{m-2})_{e}^{b}(\mu) \\
&-& {c_{m-1}(\mu-\la)\over b_{m-1}(\mu-\la)}(B^{m-2})_{a}(\mu) (T^{m-2})_{c}^{b}(\la).
\end{eqnarray*}
The indices take values in $\Asc_{m-1}$ and the level $m-2$ creation operator is $(B^{m-2})_c =
(T^{m-2})^{\ahat_{m-2}}_c$, where $\ahat_{m-2} \in \Asc_{m-2}$ is the chosen level $m-2$
pseudo-vacuum state. Just as in the bosonic case, the level $m-1$ monodromy matrix will turn out to
be proportional to a product of permutation operators yielding the unit shift operator. Thus
$\La^{m-1}(\mu)$ will be a root of unity.

For $k=0,\ldots,n+m-2$ the Bethe equations are
\begin{eqnarray}
\prod_{y=1}^{p_{k+2}} {a_{k+1}(\la_y^{k+2}-\la^{k+1}_z)\over b(\la_y^{k+2} - \la^{k+1}_z)}
\mathop{\prod_{y=1}^{p_{k+1}}}_{y\neq z} {a_{k+1}(\la^{k+1}_z-\la_y^{k+1})
b(\la_y^{k+1}-\la^{k+1}_z) 
\over a_k(\la_y^{k+1} - \la^{k+1}_z) b(\la^{k+1}_z-\la_y^{k+1})}
\prod_{x=1}^{p_k} {b(\la^{k+1}_z - \la^k_x)\over a_k(\la^{k+1}_z-\la_x^k)}
=1, \nn\\
\end{eqnarray}
where we have used the explicit formulae for the functions $a_I(\mu)$ and $b(\mu)$ to cancel off
some signs and simplify the expression a little. Notice that these are essentially the same as the
Bethe equations for the bosonic case with $a_I(0) = 1$.

The spin chain Hamiltonian for these models is the logarithmic derivative of the transfer matrix with respect to the spectral parameter $\la$. The calculation is simpler when one assumes that all the level-0 inhomogeneities vanish since the Hamiltonian is then calculable as the derivative of the $R$-matrix. Using the explicit $R$-matrix given by the coefficient functions (\ref{afn}) -- (\ref{dfn}), we have
\begin{eqnarray}
H &=& \OP{P} {d R(\la) \over d\la}\Big|_{\la \to 0} \nn\\
&=& -2\pr{1+q^2 \over 1-q^2} \mathop{\sum_I}_{|I| = 1} \sum_{a_I,b_I \in \Asc_I} e_{a_I}{}^{b_I} \otimes_s e_{b_I}{}^{a_I}
 +  \pr{2q      \over 1-q^2} \sum_{I\neq J} \mathop{\sum_{a_I \in \Asc_I}}_{b_J \in \Asc_J} (-)^{|I||J|} e_{a_I}{}^{b_J} \otimes_s e_{b_J}{}^{a_I} \nn\\
& & + \pr{2q^2    \over 1-q^2} \sum_{I < J}   \mathop{\sum_{a_I \in \Asc_I}}_{b_J \in \Asc_J} e_{a_I}{}^{a_I} \otimes_s e_{b_J}{}^{b_J}
 +  \pr{2       \over 1-q^2} \sum_{I > J}   \mathop{\sum_{a_I \in \Asc_I}}_{b_J \in \Asc_J} e_{a_I}{}^{a_I} \otimes_s e_{b_J}{}^{b_J}
\end{eqnarray}
where $\OP{P}$ is the graded permutation operator
\begin{equation}
\OP{P} = \sum_{I,\,J = 1}^{m+n-1} \mathop{\sum_{a_I \in \Asc_I}}_{b_J \in \Asc_J} (-)^{|I||J|} e_{a_I}{}^{b_J} \otimes_s e_{b_J}{}^{a_I}.
\end{equation}
The results of the algebraic Bethe ansatz allow us to compute the energy spectrum corresponding to this Hamiltonian. If we make the drastic simplification that $p_0 > p_k$ for all $k > 0$, then we can arrive at a closed form expression. By taking the logarithmic derivative of $\La^0(\la)$ we find that if the level-1 states are bosonic $|k=1| = 0$, then
\begin{equation}
E = \sum_{y = 1}^{p_1} {\sinh \gamma \over \sinh\pr{\la^1_y + \gamma} \sinh \la^1_y},
\end{equation}
where we have let $q = e^{-\gamma}$. The answer differs a bit if the level-1 states are fermionic $|k=1| = 1$:
\begin{equation}
E = p_0 - p_1 - p_0 {\cosh \gamma \over \sinh \gamma} + \sum_{y=1}^{p_1} {\cosh \la^1_y \over \sinh \la^1_y} .
\end{equation}

\section{Conclusions}
In this paper we extended the multiplicity $A_m$ models of \cite{AmModels} to the supersymmetric case. We used the nested algebraic Bethe ansatz to find the eigenvalues of the transfer matrix and the corresponding Bethe equations. Using these results, we computed the Hamiltonian and its energy levels for the specific multiplicity model associated with the supersymmetric XXZ model.
 
While the multiplicity of the model can be hidden notationally in the description of the model's $R$-matrix, the addition of multiplicity is reflected in the form of the nested ABA equations. The final level of the nesting becomes more non-trivial, as in (\ref{UnitShift}), and there are also additional terms proportional to $\delta_{p_k p_{k+1}}$ at every other level. The multiplicity also contributes to the energies in the corresponding spin chain when $p_0 = p_1$ or (some of) the level-0 inhomogeneities are non-zero.

Since the XX model is a building block for more complicated theories, notably the Hubbard model, it would be interesting to investigate theories constructed by coupling together supersymmetric multiplicity models. Generalizing the Hubbard model while maintaining integrability has proven difficult \cite{CPH, BeisertSU22, EKS, BGLZ, DFFR, MartinsHubbard}. We hope that the work presented here will shed some light on this problem.

\bigskip
\begin{center}
\noindent{\bf Acknowledgements}
\end{center}

We are grateful to Jonathan Evans for helpful conversations. 
The research of C.A.S.Y. was funded by the Leverhulme trust.

\end{document}